\documentstyle[aps,preprint]{revtex}

\preprint{\hfil{NCL96--TP5}} 

\begin{document}
\draft
\title{The Phase Transition of the Two Higgs extension of the Standard Model}
\author{Vasilios Zarikas}
\address{Department of Physics, University of Newcastle Upon Tyne, NE1 7RU\\
U.K. }
\maketitle

\begin{abstract}
We present the analysis of the phase transition for the two Higgs
electroweak model. We have found that for a wide range of parameters the
universe first tunnels to a new intermediate phase. This feature not only is
very important by itself, but also provides the essential requirements for
producing baryon asymmetry with only small explicit $CP$-violating terms in
the two Higgs tree Lagrangian.
\end{abstract}


\section{INTRODUCTION}

The mechanism of spontaneous symmetry breaking plays a big role in the
success of the standard model of the electroweak interactions. It also leads
to the possibility of phase transitions and important cosmological
phenomena. This symmetry breaking mechanism is based ultimately on the
scalar Higgs potential.

The current experimental situation with respect to the standard electroweak
model's Higgs sector gives the possibility for modifications. A model with
two scalar boson doublets can be a sensible extension of the standard model,
and may be necessary if one requires low energy supersymmetry or sufficient
baryon asymmetry in the electroweak transition \cite{zadro}, \cite{gave}.

In the two-Higgs model the desired baryogenesis at the electroweak scale has
been achieved by the introduction of a term that breaks the $CP$ invariance
explicitly at tree level. The amount of $CP$ violation required is
uncomfortably large, particularly if the two Higgs model is the relic of a
symmetry breaking at scales above the electroweak transition, when it is
more natural that any explicit $CP$ violating terms be small \cite{come}, 
\cite{dugan}. 

Here it is reported that, using the $CP$ invariant finite temperature
effective potential, domains of the universe can first tunnel towards a new
minimum before evolving to the usual minimum. When small $CP$ violating
terms are added to the potential the $CP$ violation is amplified in the
intermediate phase. This happens for a significant range of parameters that
cover also the minimal supersymmetric model. The result is that we expect
reasonably large baryon asymmetry even from a two Higgs scalar potential
with very small $CP$ violation at low temperatures.

\section{THE TREE POTENTIAL}

The two Higgs scalar potential can be written \cite{sher} using doublets of
hypercharge $+1$ as follows 
\begin{eqnarray}
V_{tr} &=&\mu _1^2\Phi _1^{\dagger }\Phi _1+\mu _2^2\Phi _2^{\dagger }\Phi
_2+\lambda _1(\Phi _1^{\dagger }\Phi _1)^2+\lambda _2(\Phi _2^{\dagger }\Phi
_2)^2+  \nonumber \\
&&\ \lambda _3(\Phi _1^{\dagger }\Phi _1)(\Phi _2^{\dagger }\Phi _2)+\lambda
_4(\Phi _1^{\dagger }\Phi _2)(\Phi _2^{\dagger }\Phi _1)+{\frac 12}\lambda
_5[(\Phi _1^{\dagger }\Phi _2)^2+(\Phi _2^{\dagger }\Phi _1)^2]
\end{eqnarray}
where $\lambda _i$ real numbers and 
\begin{equation}
\Phi _1={\frac 1{\sqrt{2}}}\left( {%
{\phi _1+i\phi _2 \atop \phi _3+i\phi _4}
}\right) ,\;\;\;\Phi _2={\frac 1{\sqrt{2}}}\left( {%
{\phi _5+i\phi _6 \atop \phi _7+i\phi _8}
}\right)
\end{equation}
The above form is the most general one satisfying the following discrete
symmetry, 
\begin{equation}
\Phi _2\rightarrow -\Phi _2,\;\;\Phi _1\rightarrow \Phi
_1,\;\;d_R^i\rightarrow -d_R^i,\;\;u_R^i\rightarrow u_R^i
\end{equation}
where $u_R^i$ and $d_R^i$ represent the charge ${\frac 23}$ and $-{\frac 13}$
right-handed quarks in the weak eigenstates. This symmetry forces all the
quarks of a given charge to interact with only one doublet and thus avoids
flavour-changing neutral currents.

If we stick to this potential there is no $CP$ violation in the symmetric
phase. It is possible however to introduce ``soft'' discrete symmetry
breaking terms of the form 
\begin{equation}
\Delta V={\rm Re}\{2\mu _3^2\Phi _1^{\dagger }\Phi _2+\lambda _6(\Phi
_1^{\dagger }\Phi _1)(\Phi _1^{\dagger }\Phi _2)+\lambda _7(\Phi _2^{\dagger
}\Phi _2)(\Phi _1^{\dagger }\Phi _2)\}
\end{equation}
The parameters $\mu _3$, $\lambda _6$ and $\lambda _7$ can be complex
numbers, providing explicit $CP$ violation at tree level through the
developed phase between the two VEV's. Although their presence will be
crucial for achieving the desired baryon asymmetry, it is possible to ignore
them for the study of the phase transition, assuming that these terms are
small.

We can perform an $SU(2)$ rotation that puts the VEV's of the field for $%
i=1,2,3$ equal to zero. Solving the system $\partial V/\partial \phi _i=0$
implies several different stationary points. One of them is the usual
asymmetric minimum that respects the $U(1)$ of electromagnetism. 
\begin{equation}
\Phi _1={\frac 1{\sqrt{2}}}\left( {%
{0 \atop u_I}
}\right) ,\;\;\Phi _2={\frac 1{\sqrt{2}}}\left( {%
{0 \atop v_I}
}\right)
\end{equation}
Another already known stationary point is the following 
\begin{equation}
\Phi _1={\frac 1{\sqrt{2}}}\left( {%
{0 \atop u_{II}}
}\right) ,\;\;\Phi _2={\frac 1{\sqrt{2}}}\left( {%
{v_{II} \atop 0}
}\right) \,e^{i\xi }  \label{sad}
\end{equation}
where $u_I,v_I,u_{II},v_{II}$ real numbers. In this extremum the charge
invariance is broken (the upper component of the doublet is charged).

The unknown free parameters in this model can be reduced to the five $%
\lambda _i$ and the ratio $\beta =u_I/v_I$. We have the following
constraints on the parameters ( Note that there are some differences with 
\footnote{%
between equation (7) and equation (5.25) in reference \cite{sher}.}\cite
{sher}. )

\begin{itemize}
\item  Stationary point $I$ is the minimum if and only if 
\begin{eqnarray}
\lambda _1>0,\;\;\;\lambda _2>0  \nonumber \\
\lambda _4+\lambda _5<0,\;\;\;\lambda _5<0  \nonumber \\
2\sqrt{\lambda _1\,\lambda _2}>|\lambda _3+\lambda _4+\lambda _5|  \label{e1}
\end{eqnarray}

\item  The potential is bounded below if and only if

\begin{eqnarray}
\lambda _1>0,\;\;\;\lambda _2>0,\;\;\;\ 2\sqrt{\lambda _1\,\lambda _2}%
>-\lambda _3  \nonumber \\
2\sqrt{\lambda _1\,\lambda _2}>-(\lambda _3+\lambda _4-\lambda _5)  \nonumber
\\
2\sqrt{\lambda _1\,\lambda _2}>-(\lambda _3+\lambda _4+\lambda _5)
\label{e2}
\end{eqnarray}

\item  Assuming that the potential is bounded below, the stationary point $%
II $ exists if and only if

\begin{eqnarray}
{\frac{-2\lambda _2(\lambda _4+\lambda _5)}{4\lambda _1\,\lambda _2-\lambda
_3(\lambda _3+\lambda _4+\lambda _5)}}<\beta ^2<{\frac{4\lambda _1\,\lambda
_2-\lambda _3(\lambda _3+\lambda _4+\lambda _5)}{-2\lambda _1(\lambda
_4+\lambda _5)}}  \nonumber \\
&&  \nonumber \\
4\lambda _1\lambda _2>\lambda _3(\lambda _3+\lambda _4+\lambda _5)
\label{e3}
\end{eqnarray}
\end{itemize}

Investigating the whole parameter space would be too time consuming. We
consider representative values of the coupling constants with absolute value
equal to $g^2$ of $SU(2)$ or $10^{-n}g^2$ and we take all the possible
combinations: some of them be equal to $\pm g^2$ and some equal to $\pm
10^{-n}g^2$ ($n$ a fixed positive integer).

It turns out that for all the cases with $\beta =0.1$ or $\beta =10$ and $%
\forall n$ the scalar bosons are too light. This suggests taking a range of
values near $\beta =1$. Also, we have found that even for $\beta \simeq 1$
and $n>1$ in only a few cases do the scalar bosons have masses above the
experimental lower bounds. The set of parameters given in Table \ref{param}
is allowed by constraints (\ref{e1}), (\ref{e2}), (\ref{e3}) and
experimental limits on the Higgs masses.

\section{THE PHASE TRANSITION}

In order to study the cosmological phase transition of the model we use the
finite temperature effective potential. We include one loop radiative
corrections to the tree potential using the temperature corrected fermion
and gauge boson masses ( We assume that the scalar masses do not alter the
potential significantly ). This gives the following potential, 
\begin{equation}
V_\beta =V_{tr}+{\frac 18}\,[\,\sum_i(M_A^2)_i\,+\,2\,m_t^2\,]\,T^2- {\frac 1%
{4\pi }}\,\sum_i(M_A^2)_i^{3/2}\,\,T
\end{equation}
$m_t$ is the top quark mass and $(M_A^2)_i$ are the eigenvalues of the gauge
boson mass matrix, 
\begin{equation}
(M_A^2)^{ab}=g^2\sum_{k=1}^2\Phi _k^{\dagger }\,T^aT^b\,\Phi _k
\end{equation}
The Lie algeblra matrices 
\begin{eqnarray}
T^a &=&\sigma ^a\hbox{~~~for~~~}a=1,2,3,  \nonumber \\
T^a &=&tI\hbox{~~~for~~~}a=4.
\end{eqnarray}
with $t=g^{\prime }/g$, $g^{\prime }$ is the $U(1)$ coupling constant.

Note that the top quark is coupled to the $\Phi_{1}$ doublet only. This is
an essential requirement in order to avoid flavor changing neutral currents
otherwise one must tune the Yukawa couplings.

It is necessary to explore the full range of the potential in order to have
a complete picture of the transition. We checked the shape of the potential
in every two dimensional plane passing through the origin, $\phi_{i}=
c\,\phi_{j}$ and $\phi_{k} =0$ in order to identify local and global minima.
These were verified by evaluating the first derivatives of the potential.

At very high temperatures the symmetry is restored and as the temperature
drops, we eventually start to get the first asymmetric minimum. There is a
barrier between this minimum and the symmetric one, so it seems likely to
have a first order transition.

When $\beta =1$ the first asymmetric true minimum appears in the plane where 
$\phi _3=\phi _7$ and $\phi _i=0$, and everything looks conventional. For $%
\beta <1$\thinspace the picture alters dramatically. Taking, for example,
the sixth set of parameters of Table \ref{param} and setting $\beta =0.8$,
one finds that for $T_1=259.1$ GeV the symmetric minimum ceases to be an
absolute minimum in favour of a new asymmetric minimum. To be precise, there
is a manifold of absolute minima satisfying $\phi _i=0$ for $\,i=1,...,4\;$
and $\phi _5^2+\phi _6^2+\phi _7^2+\phi _8^2=w(T)^2$.

As the temperature drops to $T_2=104$ GeV there is a new absolute minimum
\thinspace $\phi _3=u(T)$ and $\phi _7=v(T)$ and $\phi _i=0\,$ separated
from the symmetric one by a barrier. The part of the universe that had
already tunnelled in the previous minimum rolls towards this new one. The
position of this minimum as the universe cools shifts to the zero
temperature one.

In fact one can find an $SU(2)$ gauge that simplifies the picture
considerably. The Higgs fields tunnel first towards the new absolute minimum 
$III$, 
\begin{equation}
\Phi _1={\frac 1{\sqrt{2}}}\left( {%
{0 \atop 0}
}\right) ,\;\;\Phi _2={\frac 1{\sqrt{2}}}\left( {%
{0 \atop w(T)}
}\right)
\end{equation}
After the temperature drops further the fields roll down toward the next
absolute minimum: 
\begin{equation}
\Phi _1={\frac 1{\sqrt{2}}}\left( {%
{0 \atop u(T)}
}\right) ,\;\;\Phi _2={\frac 1{\sqrt{2}}}\left( {%
{0 \atop v(T)}
}\right)
\end{equation}
finally reaching the absolute minimum $I$, equation (4).

The matrix, 
\begin{equation}
(M_S^2)_{ij}={\frac{\partial ^2V_\beta }{\partial \phi _i\partial \phi _j}}
\end{equation}
of second derivatives of the potential can be found analytically. Figure \ref
{eig} shows the evolution of the eigenvalues of $M_S^2$ evaluated
numerically at point $III$. One of them starts positive and, as temperature
drops, becomes negative. The expectation values plotted in Figure \ref{vac}
show that the initial phase transition towards the minimum is a first order
one, while the following is a second order transition.

One can not gauge transform minimum $III$ to a minimum of the form of the
stationary point $II$, equation (\ref{sad}). Point $II$ is not a minimum at
high temperatures, even though the parameters we have taken fulfill
expression (\ref{e3}) and $II$ is an extremum at zero temperature. Also, for
completeness, it is worth to mentioning that for $\beta >1$ the universe
first tunnels towards the following absolute minimum $\phi _i=0$ for $i\neq
3 $ and $\phi _3\neq 0$ and eventually rests at the usual zero temperature
minimum.

It is well known that when a discrete symmetry is broken during a
cosmological phase transition it will produce stable domain walls. This
would be a problem because the discrete symmetry $\Phi \to -\Phi $ is broken
at minimum $I$ but unbroken at minimum $III$. This problem is overcome with
small additional terms, $\Delta V$ which break the discrete symmetry and
provide at the same time the explicit $CP$ violation.

The value of $\Phi _1$ at minimum $III$ is shifted to non--zero values by $%
\Delta V$, 
\begin{equation}
\Phi _1=\left( \matrix{0\cr z(T)\cr}\right) e^{i\theta }.
\end{equation}
The angle $\theta $ is a source of $CP$ violation in the fermion
interactions. It depends strongly on the phases of the extra terms, 
\begin{equation}
\theta \approx {\frac{(M_S^2)_{33}}{(M_S^2)_{44}}}\hbox{arg}(\mu _3^2+ \frac{%
w^2}4\lambda _7).
\end{equation}
It does not depend strongly on the magnitudes of the parameters in $\Delta V$%
. At minimum $I$ however, 
\begin{equation}
\theta \approx \case1/4(\mu _3^2+\lambda _5uv)^{-1}\hbox{Im}\left( \mu
_3^2+\lambda _6u^2+\lambda _7v^2\right) .
\end{equation}
The $CP$ violation in the lowest temperature phase is determined not just by
the arguments of the extra terms, but also by the ratios $|\lambda
_7/\lambda _5|$ etc., which can be small.

\section{Conclusions}

Kuzmin, Rubakov and Shaposhnikov \cite{krs} have argued that anomalous
baryon number violation in the electroweak interactions can explain the
origin of the cosmological baryon asymmetry. The problem is that in the
standard model the Kobayashi-Maskawa phases give too little $CP$ violation
to explain the observed baryon asymmetry \cite{gave}. In a two Higgs
extension of the standard model, an extra source of $CP$ violation can come
directly from a $CP$ non-invariant scalar potential \cite{zadro}, \cite{ckn}%
, \cite{tur}, \cite{cline}. The amount of $CP$ violation necessary for the
baryon asymmetry to entropy ratio infered from nucleosynthesis is difficult
to reconcile with tests of the electroweak model. The present work relaxes
the $CP$ requirement making more plausible the explanation of the
baryogenesis coming from a realistic model.

What we have found is that if $\beta <1$ regions of the universe first
tunnels towards a minimum in the Higgs potential which is very sensitive to
small $CP$ violating terms in the potential. The bubbles of the new phase
have significant $CP$ violation depending only on the phase of the extra
terms in the potential.

For the non-local mechanisms of baryogenesis which seem to work efficiently,
the fermions as they reflect from the bubble wall experience a space
dependent phase. The result is that the right-handed fermion particles have
different reflection coefficients from the left-handed anti-particles. From
ref \cite{tur}, this leads to the baryon asymmetry which for small
velocities $v_w$ of the bubble walls is given by 
\begin{equation}
{\frac{n_B}s}\approx {\frac{15}{2g_s\pi ^4}}v_wf^2\left( {\frac m{T_c}}%
\right) \theta
\end{equation}
for Yukawa coupling $f$ and $g_s$ spin states.

The issue of how thin the walls are depends upon the details of the
transition. A handy feature of the two Higgs model is that it can give a
stronger first order phase transition and thus thinner walls. In this case
the theory of tunneling rates using the finite temperature effective
potential up to one loop order \cite{zar} suffices and there is no need for
out of equilibrium techniques.

\newpage

\acknowledgments 

I am grateful to Ian Moss for several enlightening discussions and for
suggesting I study this model. I wish to acknowledge the support of the
State's Scholarship Foundation of Greece.

\begin{table}[tbp]
\begin{tabular}{||llllll||}
$\lambda _{1}$ & $\lambda _{2}$ & $\lambda _{3}$ & $\lambda_{4}$ & $%
\lambda_{5}$ & $\beta$ \\ \hline\hline
$g^2$ & $g^2$ & $g^2$ & $-g^2$ & $-g^2$ & $\simeq 1$ \\ \hline
$g^2$ & $g^2$ & $-0.1g^2$ & $-g^2$ & $-0.1g^2$ & $\simeq 1$ \\ \hline
$g^2$ & $g^2$ & $0.1g^2$ & $-g^2$ & $-0.1g^2$ & $\simeq 1$ \\ \hline
$g^2$ & $g^2$ & $g^2$ & $-g^2$ & $-0.1g^2$ & $\simeq 1$ \\ \hline
$g^2$ & $g^2$ & $-0.1g^2$ & $0.1g^2$ & $-g^2$ & $\simeq 1$ \\ \hline
$g^2$ & $g^2$ & $0.1g^2$ & $0.1g^2$ & $-g^2$ & $\simeq 1$ \\ \hline
$g^2$ & $g^2$ & $g^2$ & $0.1g^2$ & $-g^2$ & $\simeq 1$ \\ \hline
\end{tabular}
\par
\vspace{4mm}
\caption{Sample of allowed parameters from the various theoretical and
experimental constraints}
\label{param}
\end{table}

\begin{figure}[tbp]
\caption{ The evolution of the non zero eigenvalues of the second derivative
of the potential, evaluated at point $III$, as a function of the
temperature. }
\label{eig}
\end{figure}

\begin{figure}[tbp]
\caption{ The vacuum expectation values described in the text are plotted as
a function of the temperature.}
\label{vac}
\end{figure}

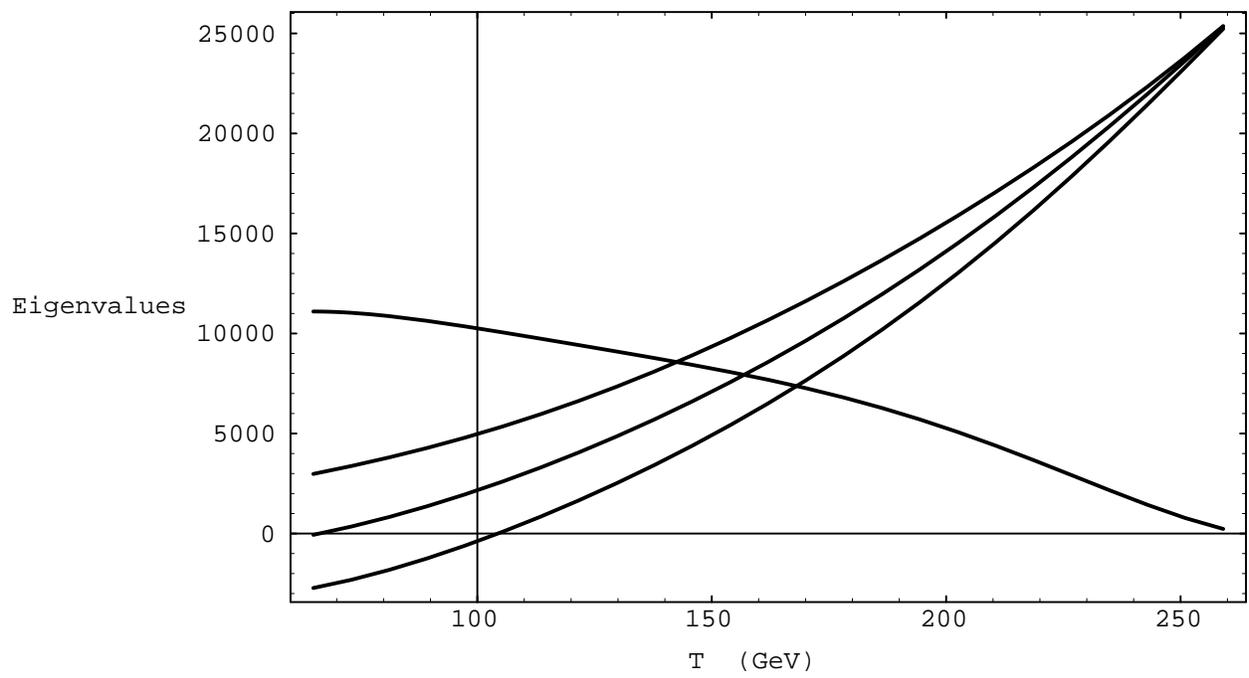
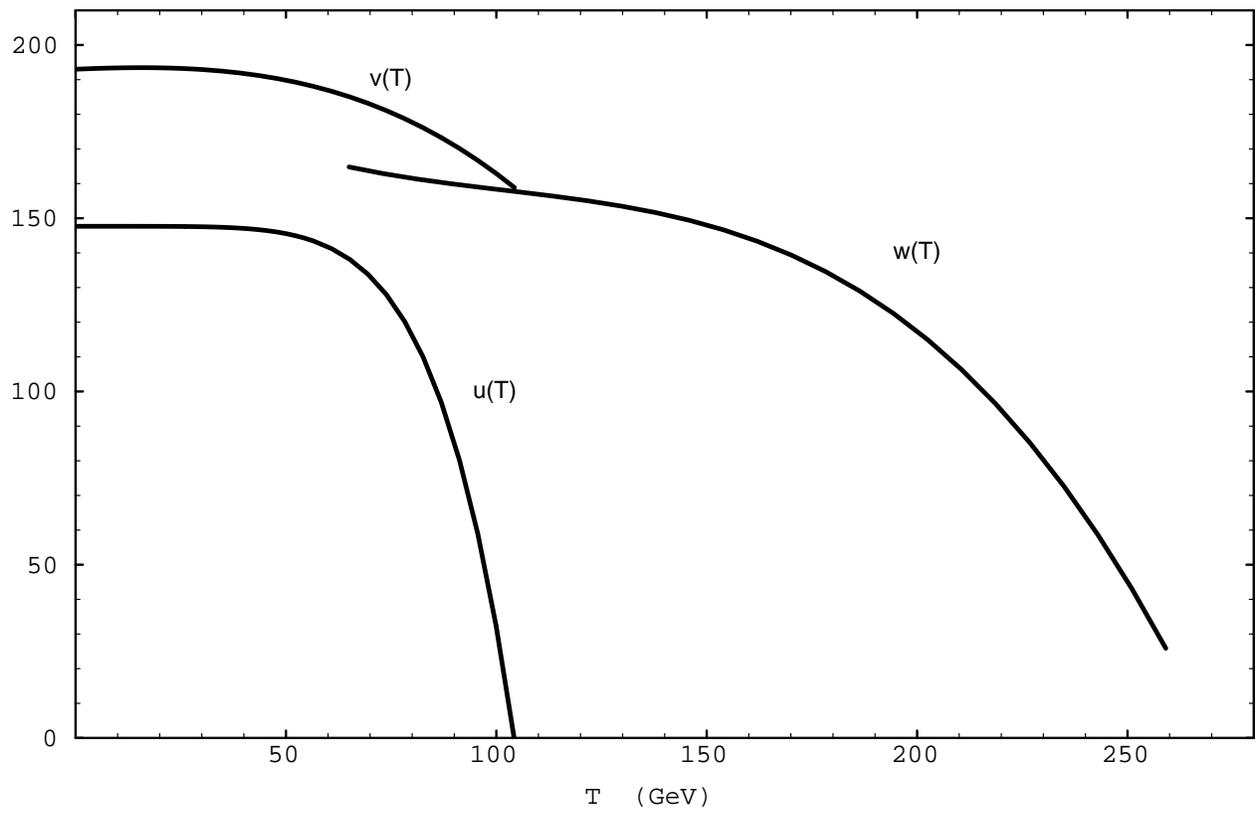
\end{document}